\def\sz{\sigma_z}
\def\sx{\sigma_x}
\def\sy{\sigma_y}
\def\s0{\hat{\bf 1}}
\def\H{{\cal H}}
\def\a+{a^{\dagger}}
\def\e{\varepsilon}
\def\D{\Delta}
\def\O{\Omega}
\def\oT{\omega_0}
\def\lx{\lambda_x}
\def\lz{\lambda_z}
\def\HRWA{{\cal H}^{RWA}}

\def\VV{\cal V}
\def\V+{{\cal V}^{\dagger}}
\def\UU{{\cal U}}
\def\U+{{\cal U}^{\dagger}}

\def\W+{{\cal W}^{\dagger}}
\def\>{\left.\right>}
\def\<{\left<\right.}
\def\){\left.\right)}
\def\({\left(\right.}
\def\d{\delta}
\def\r{\rho}
\def\b{\beta}
\def\l{\lambda}

\documentclass[prl,twocolumn,showpacs]{revtex4}
\usepackage{graphicx}
\usepackage{dcolumn}
\usepackage{bm}

\begin{document}
\title{Selective Amplification of a Quantum State}

\author{A.M. Zagoskin}
\affiliation{D-Wave Systems Inc., 320-1985 W. Broadway, Vancouver, B.C., V6J 4Y3 Canada}
\affiliation{Physics and Astronomy Dept., The University of British Columbia,
6224 Agricultural Rd., Vancouver, B.C., V6T 1Z1 Canada} 
\author{M. Grajcar}
\affiliation{Department of Solid State Physics, Comenius University, SK-84248 Bratislava, Slovakia}
\affiliation{Institute for Physical High Technology, P.O. Box 100239, D-07702 Jena, Germany}
\author{A.N. Omelyanchouk}
\affiliation{B.I.Verkin Institute for Low Temperature Physics and Engineering, Ukrainian National Academy of Sciences, Lenin Ave. 47, 310164 Kharkov, Ukraine}

\begin{abstract}
We predict a novel effect in a quantum two-level system (TLS) coupled to a
resonant cavity.  By bringing the TLS in and out of resonance with the cavity by
a series of $N$ rectangular bias pulses (the length of the $m$th pulse scaling
as $1/\sqrt{m}$), we will coherently excite the $N$-photon state, $|N\>,$ of the
cavity only if the TLS was initially in an appropriate quantum state ("go"
state). Otherwise the number of photons in the cavity will remain small compared
to $N$ ({\em selective amplification}). If the TLS was in a coherent
superposition of the "go" and "no go" states, the cavity will be in a
superposition of states, in which the state $|N\>$ will enter with the same
weight as the initial "go" component. The effect is due to
$\sqrt{N+1}$-dependence of the Rabi oscillation frequency on the number $N$ of
photons in the cavity. It is stable with respect to noise, pulse shape, finite
temperature, TLS decoherence and TLS detuning from resonance with the cavity.
The effect can be used as a means to read out a quantum state of a qubit coupled
to a resonator.
\end{abstract}
\maketitle

\section{Introduction}

A two-level system (TLS) coupled to an oscillator is a textbook example in
quantum optics \cite{Orszag}, where dressed states, Rabi oscillations and
entanglement between an atomic degree of freedom and cavity modes can be
investigated \cite{CAVITY MODES?}. More recently, this model was used to describe
the behaviour of superconducting quantum bits interacting through some kind of a
resonator or a bus \cite{Yukon?,AAA,SZ,Plastina,Yale}. Drawing the analogy with
cavity QED \cite{CAVITY MODES?}, one can entangle qubits with each other by
consecutive coupling them to the resonator, creating in the process a coherent
superposition of number states in the resonator, e.g. $|n\>$ and $|m\>$, with
close values of $n, m$.

If we could put the resonator in a superposition state $\alpha|N\>+\beta|M\>$,
with $\alpha$ and $\beta$ depending on the state of a TLS coupled to it,  the
latter can be determined by measuring the resonator,  if only $n$ is above the
detection threshold, $m$ is below it, and the difference $|m-n|$ exceeds the
noise level. We show that such a phenomenon actually takes place, under the
condition, that the energy levels of the TLS can be made (almost) degenerate,
with tunneling splitting $\D \ll \oT$, the frequency of the
resonator \footnotemark[1].\footnotetext[1]{This amplification scheme for
detection of a qubit state was first suggested by the authors in
Ref.~\onlinecite{Amsterdam}. Similar technique was proposed in Ref.~\onlinecite{Nori}
as a means of creating arbitrary Fock states in a microcavity.}

The Hamiltonian of the system is
\begin{equation}
\H = \oT \a+ a + \H _q + \H _{\rm int}.
\end {equation}
Here the TLS Hamiltonian is \begin{equation}
\H _q = -\frac{1}{2}\e\sz + \frac{1}{2}\D\sx.
\end{equation}
It is written in the basis of "physical" states, which are the eigenstates of
the qubit at infinite bias $\e$. (For a superconducting flux qubit \cite{Mooij},
which is a loop with a circulating supercurrent, such states correspond to
clockwise and counterclockwise current directions. For a charge qubit
\cite{Nakamura}, those are states of a small superconducting grain with a
definite number of Cooper pairs.) We will denote the corresponding states of the
TLS-resonator system by $|L,n\>$ and $|R,n\>$. In the same basis, the
interaction is
\begin{equation} \H _{\rm int} = (\lz \sz + \lx \sx)(\a+ +a).  
\end{equation}

In the absence of small parameters, one would be reduced to solving numerically
the equation of motion for the density matrix of the system,
\begin{equation}\label{eq_rho}
i\dot{\r} = \left[\H ,\r\right]
\end{equation}
(for the time intervals less than the dephasing and decay time in the system,
which we assume to be much longer than other time scales, $\hbar/\oT,
\hbar/\D$). We will later present the numerical solution of this equation in the
presence of noise, to demonstrate the stability of the effect.  But first let us
use the rotating wave approximation (RWA) \cite{Orszag}, which provides a clear
physical picture of the process and turns out to be quantitatively accurate for
the relevant choice of parameters.

\begin{figure}
\includegraphics[width=8cm]{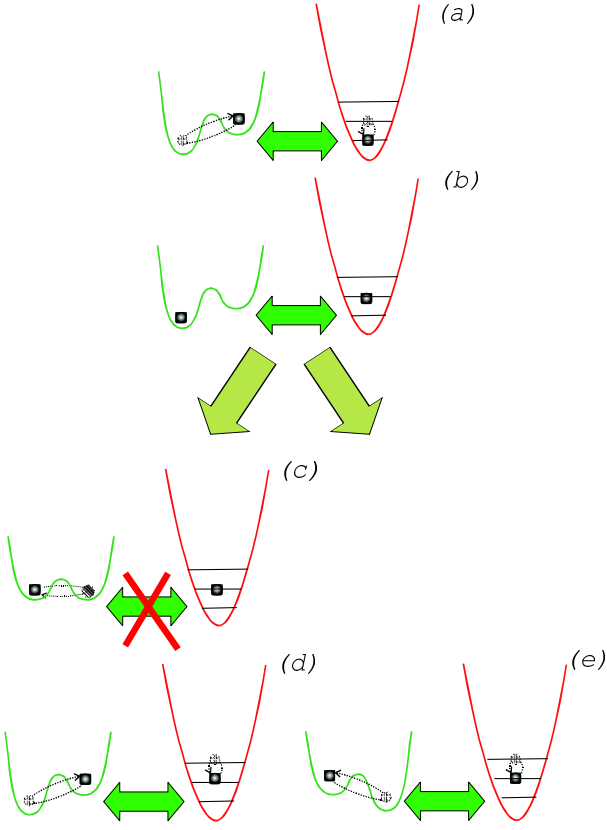}	
\caption{Physical picture of selective amplification}\label{fig1}
\end{figure}

\section{Selective Amplification in the Rotating Wave Approximation}

The condition $\D \ll \oT$ allows us to consider two separate RWA regimes.  In
the small bias regime, $\e \ll \oT,$  the TLS is effectively decoupled from the
resonator due to the large discrepancy in the relevant frequencies
(Fig.\ref{fig1}c):

\begin{equation}
\HRWA _0 = \oT \a+ a -\frac{1}{2}\e\sz + \frac{1}{2}\D\sx.
\end{equation}

In the near-resonant regime, $\O (\e) \equiv \sqrt{\D^2+\e^2} \approx \oT$, 
the system is described by the Jaynes-Cummings Hamiltonian\cite{Orszag} (Fig.\ref{fig1}a,d,e):
\begin{equation}
\HRWA _\e = \V+ (\e) \tilde{\cal H}^{RWA}_\e \VV (\e);
\end{equation}
\begin{equation}
 \tilde{\cal H}^{RWA}_\e \equiv     \omega_T a^{\dagger} a + \left[ - \frac{1}{2} \Omega(\varepsilon) \sz + g(\e) (a^{\dagger} \sigma_- + a \sigma_+) \right].
\end{equation}
Here the Pauli matrices $\sx, \sz, \sigma_{\pm} = \sx \mp i\sy$ are in the
eigenbasis of biased qubit (so that the states of the system are now $|0,n\>_\e$
and $|1,n\>_\e$). The subscripts remind, that the basis is bias-dependent;
obviously, we can make a convention  that $|0,n\>_{\infty}=|R,n\>,
|1,n\>_{\infty}=|L,n\>$ (and the opposite when $\e=-\infty$). The effective coupling is
\begin{equation}
g(\e) = \frac{\lz\D + \lx\e}{\O(\e)}.
\end{equation}
The block-diagonal
matrix $\VV(\e) = \bigoplus{\rm v}(\e)$ realizes the transformation
between this "energy" and the initial "physical" basis:
\begin{eqnarray}
{\rm v}(\e) = \frac{1}{\sqrt{2}}\left(\begin{array}{ll}
\sqrt{1+\e/\O(\e)} & -\sqrt{1-\e/\O(\e)}\\
\sqrt{1-\e/\O(\e)} & \sqrt{1+\e/\O(\e)}
\end{array}
 \right).
\end{eqnarray}

The Hamiltonian $\tilde{\cal H}^{RWA}_\e$ can be diagonalized
(e.g. \cite{Orszag}). Its eigenstates are superpositions of states with the same
excitation number (e.g., $|0,n\>_\e$ and $|1,n-1\>_\e$), and if initialized in
the state with a fixed photon number in the resonator, the system will undergo
quantum beats with frequency, dependent on $n$, which can be considered as Rabi
oscillations of the TLS in the presence of the resonator field (see
Fig.~\ref{fig1}a,d,e).  To be precise, the evolution operator,
$\exp[-it\tilde{\cal H}^{RWA}_\e],$ up to a phase factor, is given by a
block-diagonal matrix
\begin{eqnarray}
{\cal S}(\e,t) =  \left(
\begin{array}{llll}
\exp[-\frac{i}{2}t\d(\e)] & & & \\
 &  {\rm s}_0(\e,t) & & \\
 &  &  {\rm  s}_1(\e,t)& \\
 &  &  & \ddots \\
\end{array}
\right).
\end{eqnarray}
Here $\d(\e) \equiv \oT - \O (\e)$ is the detuning between the TLS and the resonator. 
The $2\times 2$ matrices ${\rm  s}_j, j=0,1,\dots$: 
\begin{widetext}
\begin{eqnarray}\label{eq_Smatrix}
{\rm  s}_j = e^{-ij\oT t}\!\!\left(\begin{array}{ll}
  \cos\left(\frac{Q(j,\e)t}{2}\right) + \frac{i\d(\e)}{Q(j,\e)} \sin\left(\frac{Q(j,\e)t}{2}\right) &
 \frac{-2ig(\e)\sqrt{j+1}}{Q(j,\e)} \sin\left(\frac{Q(j,\e)t}{2}\right) \\
 \frac{-2ig(\e)\sqrt{j+1}}{Q(j,\e)} \sin\left(\frac{Q(j,\e)t}{2}\right) &
 \cos\left(\frac{Q(j,\e)t}{2}\right) - \frac{i\d(\e)}{Q(j,\e)} \sin\left(\frac{Q(j,\e)t}{2}\right)\end{array}
 \right) 
\!\! \equiv\! e^{-ij\oT t}\!\!\left(\begin{array}{l}
 C_j(t) \: -iS_j(t) \\
 -iS_j(t) \: C_j^*(t)
 \end{array}\right)
\end{eqnarray}
\end{widetext}
describe the quantum beats between the states $|1,j\>_\e$ and $|0,j+1\>_\e$,
belonging to the same $(j+1)$th sector (labeled by the total number of
excitations in the system). The corresponding frequency is $Q(j,\e)/2$:
\begin{equation}
Q(j,\e) = \sqrt{\d^2(\e) + 4 g^2(\e)(j+1)}.
\end{equation}
The vacuum state, $|0,0\>_\e$, is in a sector of its own
and acquires only a phase factor.
 
\begin{figure}
\includegraphics[width=8cm]{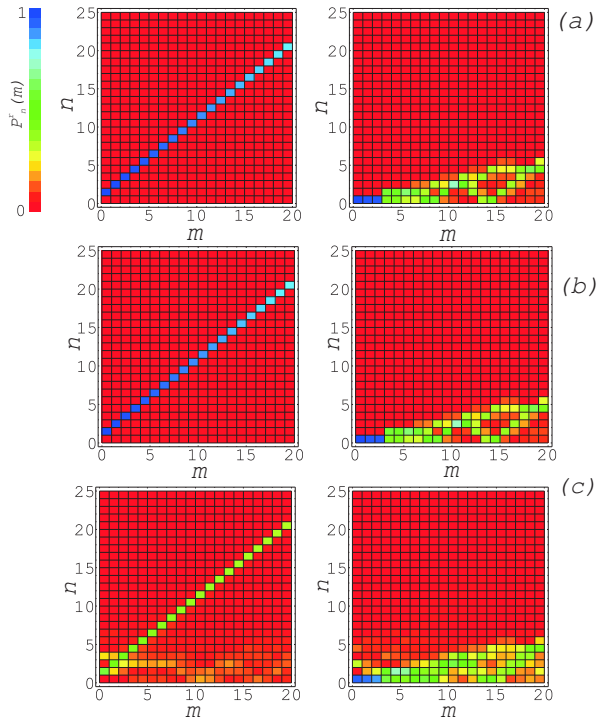}
\caption{Probability distribution of occupation numbers of the resonator, $P_n(m) \equiv \rho^{r}_{nn}(m),$ as a function of number of bias cycles, $m$, calculated using RWA. In the units of $\Delta$, the parameters are $\omega=10$ (resonator frequency), $\l_z=0, \l_x=0.8$ (TLS-resonator couplings), $\delta=0$ (perfect resonance), $|\e|=9.9499$. The length of the $m$th bias pulse is $3 T_0/\sqrt{m}$, where $T_0 = 5.9202$ is the Rabi half-period for empty resonator.
(a) Zero temperature: empty resonator; initial density matrix of the TLS is $\rho^{\rm TLS}(0) = |1\>_\e\<1|_\e$ ("go" state, left), $|0\>_\e\<0|_\e$ ("no go" state,right). 
(b) Same for finite temperature (resonator in equilibrium at $T=1$); 
(c) Same for $T=20$.
}\label{fig2}
\end{figure}

Now we can discuss qualitatively the evolution of TLS and resonator.  Suppose
the system is initially in a pure state $$|\Psi(0)\> =
\sum_{j=0}^{\infty}\sum_{\alpha=0}^{1}C_{\alpha j}|\alpha,j\>, \sum_{\alpha,j}
|C_{\alpha j}|^2=1$$ and  assume for simplicity exact resonance, $\d(\e)=0$. It
is convenient to introduce complex two-dimensional vectors ${\bf C}_{j}^{\rm T}
= (C_{0 j}, C_{1,j-1})$.   Then, according to  (\ref{eq_Smatrix}), after time
$T_p = \pi/Q(p-1,\e)$ the vector ${\bf C}_p$ flips (up to a total phase factor),
while  all other vectors ${\bf C}_{l\neq p}$ turn by some angle determined by
the ratio $T_p/T_l$. If wait for a large odd multiple of $T_p$, the rotation
angles of all $l\neq p$-vectors become quasirandom. This will be important for
the following. Still, the maximum number of photons in the reservoir cannot
change by more than one. (In the simplest case, we transformed $|1,p-1\> \to
|0,p\>$.) 

In other words, we coherently discharged the TLS into the resonator
(Fig.~\ref{fig1}b).  In order to pump more photons, we must recharge the TLS by
switching {\em between} the excitation-number sectors  (e.g. $|0,p\> \to
|1,p\>$). 

There are two possibilities. We can suddenly lift the bias, $\e \to 0$,
decoupling the resonator from the TLS (Fig.~\ref{fig1}c) (as suggested in
Refs.~\onlinecite{Amsterdam,Nori}). The TLS state $|0\>_\e \approx |R\>$ is a superposition
of the eigenstates of unbiased TLS,     $|0(1)\>_0 = (|R\>+(-)|L\>)/\sqrt{2}.$
After an odd multiple of $T_0 = \pi/\D$, it will approximately become the
$|L\>$-state, and if then we reapply the same bias $\e$, the system will find
itself in the desired state $|1,p\>_\e$ (Fig.\ref{fig1}d). The accuracy of this
transformation is easily calculated. 

\begin{figure}
\includegraphics[width=8cm]{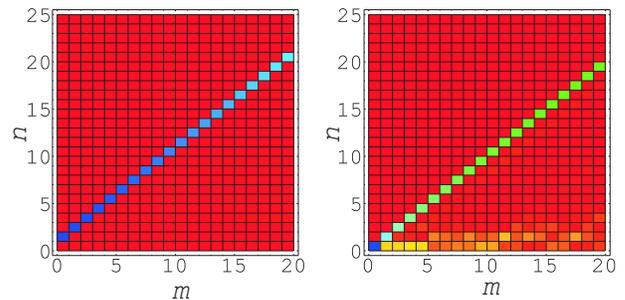}
\caption{Importance of phase randomization: time dependence of $P_n$ for the same parameters as in Fig.\ref{fig2}(a), but with the bias pulse length $T_m =   T_0/\sqrt{m}$. The "no go" state is also amplified and can not be reliably distinguished from the "go" state. Further increase of the $\mu$-factor beyond 3 will not qualitatively change the outcome (not shown). }\label{fig3}
\end{figure}

\begin{figure}
\includegraphics[width=8cm]{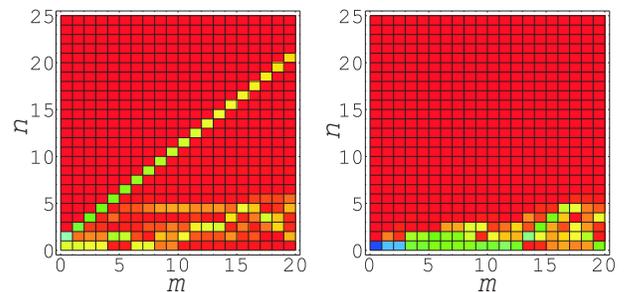}	
\caption{Role of detuning: $\omega=10$, $\l_z=0, \l_x=0.8$, $\delta=1$.}\label{fig5}
\end{figure}

Another method is instead to reverse the sign of the
bias \footnotemark[2]\footnotetext[2]{AZ is grateful to B. Wilson for pointing it
out.}, Fig.~\ref{fig1}d. Now we simply switch between the $\e$-basis and
$(-\e)$-basis, with the transformation matrix $\VV(-\e)\V+(\e) = \UU(\e) \equiv \bigoplus
{\rm u}(\e)$, where 
\begin{eqnarray}
{\rm u}(\e) = \left(\begin{array}{ll} \frac{\D}{\O(\e)} & -\frac{\e}{\O(\e)}\\
\frac{\e}{\O(\e)} & \frac{\D}{\O(\e)} \end{array}\right) \approx -i\sy
\end{eqnarray}
if $\D \ll \e,\O(\e)$, and come to the "recharged" state $|1,p\>_{-\e}$.

We will be discussing the later method, as more straightforward. Nevertheless,
if the actual physical system described by a TLS does not allow to switch the
sign of the bias, but only to bring the levels close to the degeneracy, the
former method can be applied to the same effect.

Note that in both cases the TLS-switching transformation does {\em not} depend
on the resonator occupation number. On the contrary, the transition process of
the additional photon to the resonator is sensitive to it; therefore we need now
to wait an odd multiple of $T_{p+1}$ to flip the vector ${\bf C}_{p+1}$; all
other vectors will again turn by quasirandom angles. 

Now we see how the selective amplification works. If apply to the TLS a series
of resonant bias pulses $(\e,-\e,\e,-\e,\dots)$ of duration $(\mu_1 T_p, \mu_2
T_{p+1}, \mu_3 T_{p+2}, \mu_4 T_{p+4}, \dots)$ ($\mu_j$ is an arbitrary odd
number, then the weight of the $|1,p\>$-component of the initial wave function
of the system will be transferred upwards (up to a phase factor): $C_{1p} \to
C_{0,p+1} \to C_{1,p+1} \to C_{0,p+2} \to \cdots$ On the other hand, all other
weights will be quasirandomly redistributed leading to, at best, diffusive
growth of the maximal occupation number of the resonator. If we measure this
number by a detector with  $N$-photon threshold after $N-p+s$ steps, the
probability of nonzero reading will be given by the initial weight of the state
$|1,p\>_e$, as long as the dispersion of the average photon number is less than
$s^2$. Therefore we can measure the initial quantum state of the system.


\begin{figure}
\includegraphics[width=8cm]{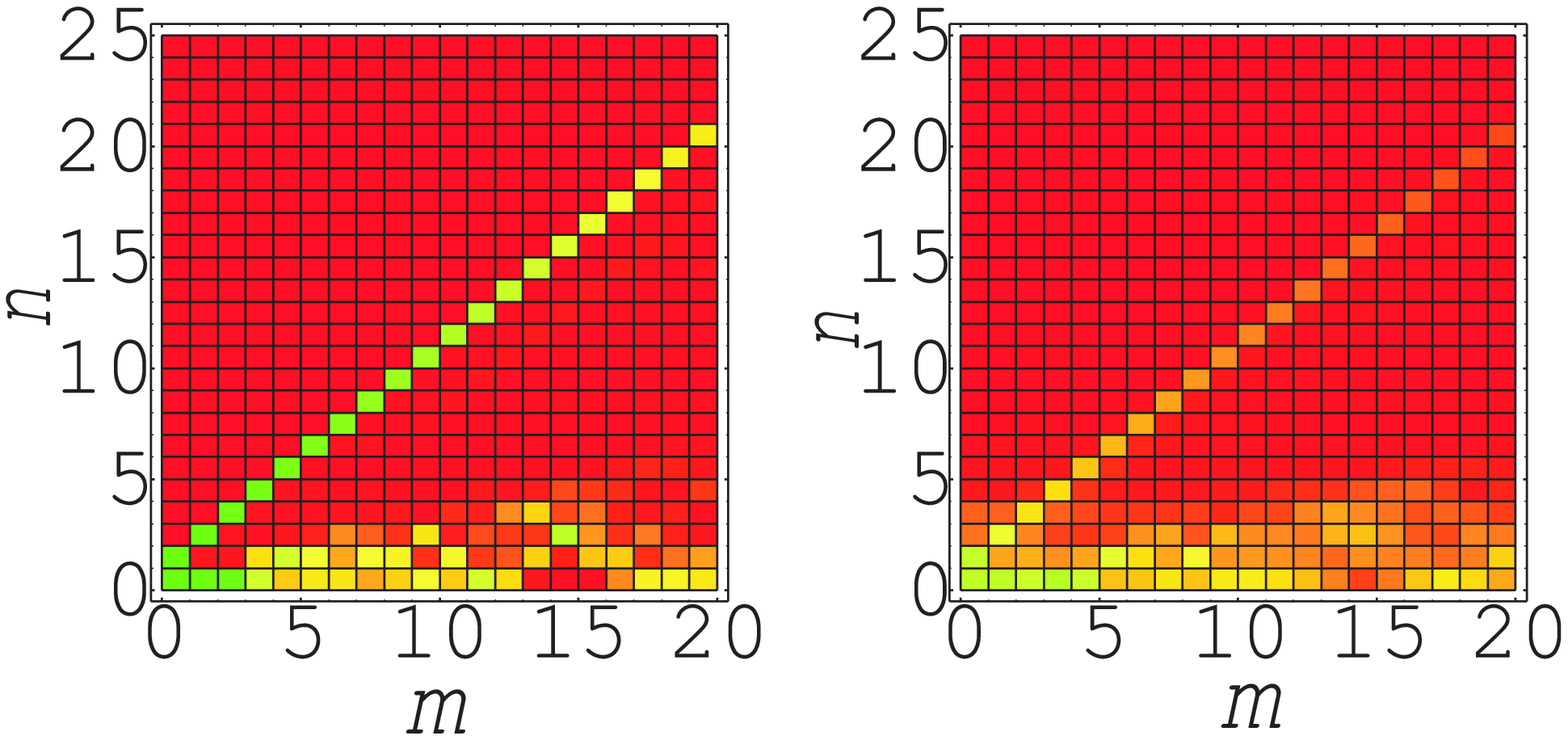}
\caption{Selective amplification of a superposition state, $\rho^{\rm TLS}(0) = \frac{1}{2}(|1\>+|0\>)_\e(\<1|+\<0|)_\e$.
The parameters are $\omega=5$, $\l_z=0, \l_x=0.1$, $\delta=0$, $|\e|=4.8990$. The length of the $m$th bias pulse is $3 T_0/\sqrt{m}$, where $T_0 = 48.0956$ is the Rabi half-period for empty resonator. Temperature $T=0; T=1$ (left), $T=10$ (right) (the pictures for $T=0$ and $T=1$ are visually indistinguishable on the chosen scale -cf. Fig.2(a,b)). Note that the effect survives in spite of smaller coupling and smaller mismatch between $\omega$ and $\D$. The state of the resonator after 20 pulses (rightmost column) is approximately $(|0\>+|20\>)/\sqrt{2}$.}\label{fig4}
\end{figure}

\begin{figure}
\includegraphics[width=8cm]{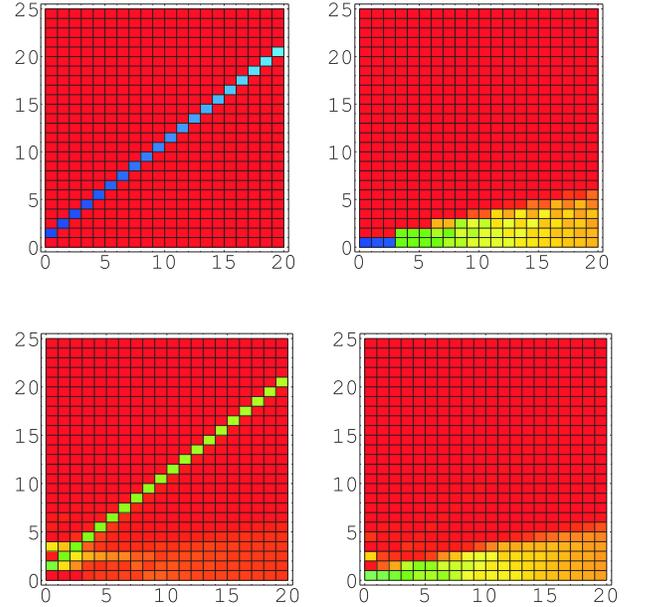}
\caption{Role of decoherence (the state of TLS is measured after every bias cycle, as described in the text): $\omega=10$, $\l_z=0, \l_x=0.8$, $\d=0$. The temperature $T=0$ (top), $T=20$ (bottom).}\label{fig6}
\end{figure}

The effects of finite temperature can be easily taken into account under a
natural assumption that the energy relaxation time for our system is much longer
than the observation time. (Otherwise it would be impossible to observe quantum
coherence at all.)  The system will be then described by the density matrix
$\r(t)$, with the initial value $\r(0) = \r^{\rm TLS}(0)\otimes\r^{r}_{eq,T}$,
where $\r^{r}_{eq,T}$ is the equilibrium density matrix of the resonator at
temperature $T$. Its time evolution is given by
\begin{equation}\label{eq_evolution1}\left(\prod {\cal
S}(\e_i,t_i-t_{i-1})\right) \r(0) \left(\prod {\cal S}(\e_i,t_i-t_{i-1})\right)^{\dagger}. 
\end{equation}

The results are presented in Figs.(\ref{fig1}-\ref{fig6}), where the diagonal
elements of the resonator density matrix after the $m$th bias step, $P_n(m)
\equiv \rho^{r}_{nn}(m),$  are plotted. Here $\r^r(t) = {\rm tr}_{TLS} \r(t)$. A
runaway increase of the resonator occupation number for the "go" state is
clearly seen. (The detailed description is given in the captions.)

The calculations were performed for different choices of parameters. In
particular, the choice for Fig.~\ref{fig4} is comparable to the parameters of a
charge qubit coupled to a strip line \cite{Yale}. 

Fig.~\ref{fig3} demonstrates the crucial importance of phase randomization: after
only half Rabi period, the "no go" state is not sufficiently wiped out by
destructive interference and also produces a significant runaway component. 

 On the other hand,  the stability of the effect with respect to finite
 temperature, detuning, and variation of system's parameters is remarkable. Note
 that the superposition initial state of the TLS, with the vacuum initial state
 of the resonator (Fig.~\ref{fig4}), leads after $N$ bias pulses to the resonator
 state of the form $\alpha |N\> + \b|M\> $ plus small admixture of other
 low-lying states.

The role of decoherence in the TLS warrants special attention. Each bias
application increases the risk of destroying its phase coherence. In order to
model this process, we repeated the above calculations, wiping out the elements
of $\r$ corresponding to coherent TLS states (in our convention, all $\r_{ij}$
with $i$ and $j$ of different parity) after each bias application. The rest
off-diagonal elements of $\r$ correspond to quantum correlations in the
resonator, which will likely have a much longer decoherence time than the TLS
\cite{AAA,Izmalkov}. The results presented in Fig.~\ref{fig6} show that there is
virtually no effect of such decoherence on the evolution of the "go" state,
while the structure of the "no go" probability amplitude is smeared out. This is
physically clear: after a bias cycle, the "go" state is factorized, and
measuring the TLS state (which was essentially done to model decoherence) does
not affect the resonator state, which keeps all the relevant quantum
correlations. For the  "wrong" states the process reduces to an additional
averaging,  wiping out the remaining structure of the density matrix.

\section{Analytical Solution}

The decoherence model used in the previous section provides a crucial simplification, which allows an analytical solution of the problem.

Let us introduce the rotated density matrix of the system after the $n$th step:
\begin{eqnarray}
\tilde{\r}(t_n) \equiv {\cal V}(\e_{n+1})\r(t_n){\cal V}^{\dagger}(\e_{n+1}).
\end{eqnarray}
Substituting here the time evolution (\ref{eq_evolution1}) of $\r$, we obtain 
\begin{eqnarray}
\tilde{\r}(t_n) = {\cal V}(\e_{n+1}){\cal V}^{\dagger}(\e_{n}){\cal S}(\e_n,t_n-t_{n-1})\times\\ \times\r(t_n){\cal S}^{\dagger}(\e_n,t_n-t_{n-1}){\cal V}(\e_{n}){\cal V}^{\dagger}(\e_{n+1}).\nonumber
\end{eqnarray}
In order to proceed, we must make some simplifications. First, assume that $e_n = -e_{n-1}$ and $|\e|\gg\D$. This replaces the products ${\cal V}(t_n){\cal V}^{\dagger}(t_{n-1})$ with simple block-diagonal $i\sz$-matrices. Second, use RWA. Third, assume that the TLS loses coherence after every cycle, i.e. keep only the elements of the desnity matrix with the indices of the same parity. Then we obtain for the relevant elements of the matrix $\tilde{\r}$ ($\mu,\l = 1,2,\dots$:
\begin{eqnarray*}
\left[\tilde{\r}(t_{n+1})\right]_{2\mu,2\l} = \nonumber\\ = \left[\tilde{\r}(t_{n})\right]_{2\mu,2\l} C_{\mu-1}(t_n)C_{\l-1}^*(t_n)e^{-i(\mu-\l)\oT(t_n-t_{n-1})} + \\
+ \left[\tilde{\r}(t_{n})\right]_{2\mu+1,2\l+1} S_{\mu-1}(t_n)S_{\l-1}(t_n)e^{-i(\mu-\l)\oT(t_n-t_{n-1})}; \nonumber \\
\left[\tilde{\r}(t_{n+1})\right]_{2\mu+1,2\l+1} = \nonumber\\ = \left[\tilde{\r}(t_{n})\right]_{2\mu+1,2\l+1} C_{\mu-1}^*(t_n)C_{\l-1}(t_n)e^{-i(\mu-\l)\oT(t_n-t_{n-1})} + \\
+ \left[\tilde{\r}(t_{n})\right]_{2\mu,2\l} S_{\mu-1}(t_n)S_{\l-1}(t_n)e^{-i(\mu-\l)\oT(t_n-t_{n-1})},\nonumber
\end{eqnarray*} 
while $\left[\tilde{\r}(t_{n+1})\right]_{11} = \left[\tilde{\r}(t_{n+1})\right]_{11}.$
Here $C_j,S_j$ are the elements of the evolution matrix (\ref{eq_Smatrix}).
Different diagonals of the density matrix thus transform independently. We are interested primarily in the main diagonal, the elements of which, ${\cal R}_j^n \equiv \tilde{\r}_{jj}(t_n),$ describe the photon numbers in the resonator (the probability for the resonator to contain $p$ photons at the moment $t_n$ is $P_n(p) = {\cal R}_{2p+1}+{\cal R}_{2p+2}$, $p=0,1,\dots$). The recursive relations for $\cal R$'s are
\begin{eqnarray}\label{eq_recursive}
\left\{
\begin{array}{l}
{\cal R}_1^{n+1} = |C_0(t_n)|^2 {\cal R}_2^n + S_0(t_n)^2 {\cal R}_3^n; \\ 
{\cal R}_2^{n+1} = {\cal R}_1^n;  \\
(p=1,2,\dots)\\
{\cal R}_{2p+1}^{n+1} = |C_p(t_n)|^2 {\cal R}_{2p+2}^n + S_p(t_n)^2 {\cal R}_{2p+3}^n; \\
{\cal R}_{2p+2}^{n+1} = |C_{p-1}(t_n)|^2 {\cal R}_{2p+1}^n + S_{p-1}(t_n)^2 {\cal R}_{2p}^n;\\
\end{array}\right..
\end{eqnarray}

They can be further simplified if take into account the quasirandomization. Indeed, the coefficients $|C_p(t_n)|^2, S_p(t_n)^2$ contain $\cos^2 (\sin^2) \left(Q_p(t_n) t_n/2\right),$ and unless $n=p$\footnotemark[3]\footnotetext[3]{Or, more generally, $n=p-q$. This would correspond to a pulse sequence, which amplifies the initial state with $q$ photons in the resonator.}, the oscillating terms will average to zero: 
\begin{eqnarray}
\<|C_p(t_n)|^2\> =  \frac{1}{2}\(1+\frac{\d^2}{Q_p(t_n)^2}\)\(1-\d_{pn}) + \\ + \left[\frac{1}{4}\<\d\phi_p^2\>
\(1-\frac{\d^2}{Q_p(t_n)^2}\)+\frac{\d^2}{Q_p(t_n)^2}\right]\d_{pn};\nonumber\\
\<S_p(t_n)^2\> =  \frac{1}{2} \frac{4(p+1)g(t_n)^2}{Q_p(t_n)^2}\(1-\d_{pn}) + \\
+\frac{1}{2} \frac{4(p+1)g(t_n)^2}{Q_p(t_n)^2}\left[1-\frac{1}{4}\<\d\phi_p^2\>\right]
\d_{pn},\nonumber
\end{eqnarray}
where we introduced the noise through the dispersion of phase gained by the "go" state on the $n$th step,$\<\d\phi_p^2\>$. The above expressions are explicitly unitary.

Now for illustration assume $\d = 0$ (exact tuning) and $\<\d\phi_p^2\> =0$ (no external noise).  Now the coefficients  $c_{np} \equiv \<|C_p(t_n)|^2\>  = (1/2)(1-\d_{pn}),$ $s_{np} \equiv (\<S_p(t_n)^2\> = (1/2)(1+\d_{pn}),$ and  defining  $c_{n,-1} = 1,$ $s_{n,-1} =0,$ we simplify the  recursion (\ref{eq_recursive}) to
\begin{eqnarray}\label{eq_recursive_simple}
\left\{
\begin{array}{l}
{\cal R}_{2p+1}^{n+1}  = c_{np}{\cal R}_{2p+2}^{n} + s_{np}{\cal R}_{2p+3}^{n};\\
{\cal R}_{2p+2}^{n+1}  = c_{n,p-1}{\cal R}_{2p+1}^{n} + s_{n,p-1}{\cal R}_{2p}^{n}; 
\end{array}\right.\\
(n,p = 0,1,\dots). \nonumber
\end{eqnarray}

First, note that there is a solution ${\cal R}_{2p+1}^n = 0, {\cal R}_{2p+2}^n = \d_{p,n}.$ It describes the amplification of the "go" state: the distribution of photon number in the resonator $P_n(p) = \d_{p,n}$. We neglected noise and detuning, therefore we did not reproduce the slow probability spread seen in numerics (cf. Fig.2); the average photon number after $n$ steps $\<p\>_n^{\rm"go"}=n.$

Second, for "no go" terms, $n \neq p, p-1,$ Eq.(\ref{eq_recursive_simple}) reduces to
\begin{eqnarray}
\left\{
\begin{array}{l}
{\cal R}_{2p+1}^{n+1}  = \frac{1}{2}{\cal R}_{2p+2}^{n} + \frac{1}{2}{\cal R}_{2p+3}^{n};\\
{\cal R}_{2p+2}^{n+1}  = \frac{1}{2}{\cal R}_{2p+1}^{n} + \frac{1}{2}{\cal R}_{2p}^{n}; 
\end{array}\right.,
\end{eqnarray}
 which yields in the continuous limit the diffusion equation for the photon number distribution function:
 
 \begin{equation}
 \partial_n P_n(p) = \frac{1}{4} \partial^2_{pp} P_n(p),
\end{equation}
with the reflective boundary condition at $p=0$. The solution for the initial condition $\delta(p-p_0),$
\begin{equation}
P_n(p|p_0) = \left[e^{-\frac{(p-p_0)^2}{n}}+e^{-\frac{(p+p_0)^2}{n}}\right]/\sqrt{\pi n},
\end{equation}
describes the diffusive spread of "no go" states seen in the 
numerics. In particular, for $p_0=0$ the average photon number after $n$ steps $\<p\>_n^{\rm"no go"}=\sqrt{n/\pi}.$

\section{Exact Numerical Results and Noise Effects}

\begin{figure}
\includegraphics[width=8cm]{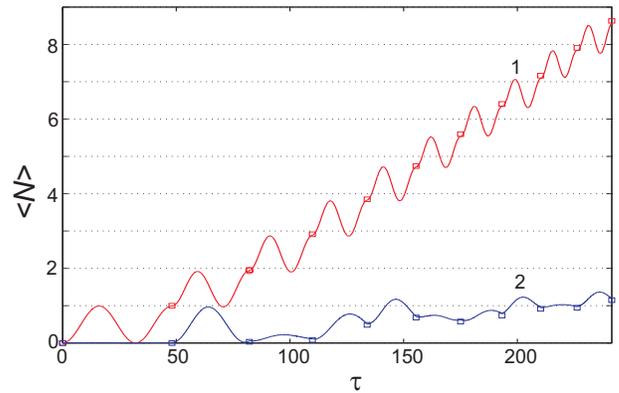}
\caption{Selective amplification: exact numerical results for $\omega=5, \lz=0, \lx=0.1, \d=0, T=0$, for "go" (curve 1) and "no go" (curve 2) states. The average resonator occupation number $\<N\>$ is plotted versus dimensionless time $\tau = \hbar/\D$. Squares show the results obtained in RWA.    }\label{fig1AO}
\end{figure}

\begin{figure}
\includegraphics[width=8cm]{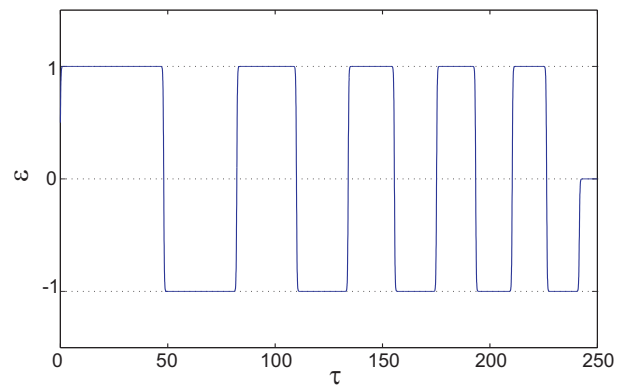}
\caption{The bias pulse sequence used in the calculations of Fig.7. The (absolute) sharpness of individual pulses is independent on the pulse duration.
   }\label{fig3AO}
\end{figure}

A direct numerical solution of the equation (\ref{eq_rho}) is also possible. In
Fig.~\ref{fig1AO} we show the resulting time evolution of the average occupation
number of the resonator, $\<N\>$, for "go" (curve 1) and "no go" (curve 2)
states at zero temperature. Unlike RWA, here we could use the realistic, smooth
pulse shape.
(Fig.~\ref{fig3AO}). 
Still, the results obtained in RWA (squares) are
remarkably accurate. 

\begin{figure}
\includegraphics[width=8cm]{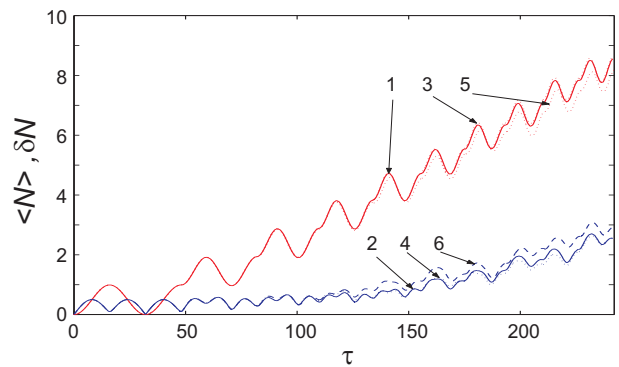}
\caption{Effect of noise on selective amplification. Equation (4) was solved with the pulse sequence of Fig.8, modified by the  Gaussian fluctuations of the pulse duration with standard deviation $\d t$ (in units $\hbar/\D$). 
The average occupation number, $\<N\>$, and r.m.s. deviation, $\d N$, are plotted for $\d t = 0$ (curves 1 and 2), $\d t = 0.01$ (curves 3 and 4), and $\d t = 0.1$ (curves 5 and 6 respectively).
 }\label{fig2AO}
\end{figure}

So far we did not discuss the effects of external noise. The most obviously
dangerous will be the fluctuations of amplitude and  duration of the bias
pulses. Their action is similar, since either primarily introduces the phase
error between the Rabi oscillations of the TLS and applied pulse sequence. We
therefore investigate the case of random pulse duration. Different models
can be considered depending on the expected source of fluctuations (e.g.
telegraph noise, 1/f noise etc). For our purposes it is enough to consider here
a Gaussian noise  with standard deviation $\d t$, most likely determined by the
properties of the pulse generator.

The influence of such noise is presented in in Fig.~\ref{fig2AO}, where both
$\<N\>$ and the r.m.s. deviation $\d N$ are plotted as functions of time for different
values of $\d t$. The effect is again stable. An obvious estimate for the
maximal number of excitations, which can be coherently pumped into the
resonator, is \begin{equation} N_{\rm max} = \left(T_0/\d t\right)^2.
\end{equation} After that the phase error introduced by the noise will be of the
same order as the phase gain during the cycle, thus destroying the constructive
interference required for the effect.

In conclusion, we predict a new effect, selective amplification of a quantum
state of a TLS, allowing a coherent pumping of excitations in a resonator by a
special sequence of bias pulses applied to the TLS.  The effect can be used in
order to read out a state of a quantum system coupled to a resonator.
The effect will be stable with respect to external noise,
finite temperature, detuning between the TLS and resonator frequency, and
decoherence in TLS.

\acknowledgements{
We are grateful to B. Wilson and J.P. Hilton  for helpful suggestions, to M.H.S.
Amin, M. Everitt, E. Il'ichev, A. Izmalkov, A. Maassen van den Brink, R.
Schoelkopf,     A. Smirnov, P. Stamp, A. Steinberg, M. Steininger, and J.Young  for many
stimulating discussions, and to I. Slobodov for technical assistance.}

\end{document}